\documentclass[runningheads]{llncs}
\usepackage{graphicx}

\usepackage{subcaption} 

\usepackage{multirow}

\usepackage{hyperref}
\makeatletter
\g@addto@macro{\UrlBreaks}{\UrlOrds}
\makeatother

\usepackage{listings}
\usepackage{xcolor}
\usepackage{xurl}

\begin{document}
\title{Where Did the Web Archive Go?}
\titlerunning{Where Did the Web Archive Go?}
%
\author{Mohamed Aturban\inst{1}\orcidID{0000-0001-7648-9082} \\ 
Michael L. Nelson\inst{2}\orcidID{0000-0003-3749-8116} \\ 
Michele C. Weigle\inst{2}\orcidID{0000-0002-2787-7166}}

\authorrunning{M. Aturban et al.}
%
\institute{Columbia College, Columbia MO 65216, USA \\
\email{maturban@ccis.edu} \and
Old Dominion University, Norfolk VA 23529 USA \\
\email{\{mln,mweigle\}@cs.odu.edu}}
\maketitle              
\begin{abstract}
To perform a longitudinal investigation of web archives and detecting
variations and changes replaying individual archived pages, or
mementos, we created a sample of 16,627 mementos from 17 public web
archives.  Over the course of our 14-month study (November, 2017
-- January, 2019), we found that four web archives changed their
base URIs and did not leave a machine-readable method of locating
their new base URIs, necessitating manual rediscovery.  Of the
1,981 mementos in our sample from these four web archives, 537 were
impacted: 517 mementos were rediscovered but with changes in their
time of archiving (or \emph{Memento-Datetime}), HTTP status code,
or the string comprising their original URI (or \emph{URI-R}), and
20 of the mementos could not be found at all.

\end{abstract}

\section{Introduction}

\setcounter{footnote}{0}
Web archives are established with the objective of providing permanent access to archived web pages, or \emph{mementos}.\footnote{This is an expanded version of a paper accepted at Theory and Practice of Digital Libraries (TPDL) 2021.}  Mementos should be accessible in web archives even after the corresponding live web page is no longer available.  The Uniform Resource Identifier (URI) \cite{rfc:3986} of the archived web page should not change over time, otherwise this defeats the purpose of using archived URIs. When web archives change their infrastructure, resulting in new base URIs for mementos, there should be machine-readable breadcrumbs left so that the older mementos still work.

We wanted to study the fixity of archived web pages, so we gathered a diverse set of mementos from 17 web archives distributed over 1996-2017. Our longitudinal experiment involved replaying the same mementos over the course of 14 months \cite{maturban_dissertation,aturban-archival-fixity,maturban2018computefixity,maturban_discovering_mementos}.  During our study, we noticed that we were no longer able to access any mementos from four web archives (Library and Archives Canada, the National Library of Ireland, the Public Record Office of Northern Ireland, and Perma.cc) at certain points, and there was no machine-readable redirection to the new URIs. This paper outlines our discovery of the disappearance of these mementos and our efforts to find their new locations.

\section{Background}
\label{sec:background}

Memento \cite{memento:rfc} is an HTTP protocol extension that allows for content-negotiation of web resources in the time dimension.  The Memento protocol is supported by most public web archives, including those included in this study. In the Memento framework, the identifier of an original resource from the live Web is a \emph{URI-R}, and the identifier of an archived version of that resource at a particular point in time is a \emph{URI-M}, or \emph{memento}.

When a request is made to a Memento-compatible web archive for a URI-M that the archive holds, the archive will include Memento headers in the HTTP response.  In particular, the Memento-Datetime HTTP Response header (e.g., \texttt{Memento-Datetime: Sun, 08 Jan 2017 09:15:41 GMT}) is sent by the archive to indicate the datetime at which the resource was archived.  

A request for a URI-M indicates both the URI-R and the Memento-Datetime requested.  If an archive does not have a memento for the requested URI-R at that particular Memento-Datetime, the archive may return an HTTP \texttt{30x Redirect} status with a \texttt{Location} response header indicating the temporally closest URI-M that the archive does have available. 

Many, though not all, Memento-compatible web archives construct URI-Ms that contain both the URI-R and the Memento-Datetime. For example, for the URI-M 
\url{http://www.collectionscanada.gc.ca/webarchives/20060208075019/http://www.cdc.gov/}, the URI-R is \url{http://www.cdc.gov/} and the Memento-Datetime is represented by the 14-digit date string \texttt{20060208075019}, which is Wed, 08 Feb 2006 07:50:19 GMT.


Web archives can differ in how they 
handle URI-Ms that returned an HTTP \texttt{404 Not Found} or \texttt{503 Service Unavailable} status code during capture. Some archives will return an HTTP \texttt{200 OK} status code and include the archived error page in the HTTP response body.  Other archives, such as the Internet Archive and Archive-It, will respond with the original status code; they return an archived \texttt{404 Not Found} for URI-Ms that returned a \texttt{404 Not Found} upon capture.

When a new archive receives an archived collection from the original archive, it may apply some post-crawling techniques to the received files (e.g., WARC files) including deduplication, spam filtering, and indexing. This may result in mementos in the new archive that have  different values of the Memento-Datetime compared to their corresponding values in the original archive.

\section{Methodology}

Our original study accessed 16,627 mementos from 17 public web archives 39 times over a period of 14 months (Nov 2017 - Jan 2019); the details of data selection are described elsewhere \cite{maturban_discovering_mementos}. For each URI-R chosen, we used the LANL Memento Aggregator \cite{bornand2016routing} in November 2017  to discover URI-Ms in different web archives. The data set is available in GitHub \cite{maturban-dataset}. Our goal was to study the fixity of mementos. We would expect that replaying the same memento over time should result in the same representation, but that is not always the case. During the time period of this study, we found instances where none of the mementos from particular archives were available. This led us to the investigations we report in this paper. 

We used the Squidwarc headless crawler \cite{squidwarc} to load each URI-M (including executing JavaScript to ensure loading all embedded resources) and download the contents into a WARC file \cite{isowarc}. Saving the data in WARC files allowed us to record all HTTP response headers and content for all of the resources that made up the \emph{composite memento} \cite{ainsworth2014framework}.

In our analysis, we refer to the archive from which mementos have moved as the \emph{original archive} and the archive to which the mementos have moved as the \emph{new archive}.  We are strict in our approach to determine if a memento in the new archive is the same as a corresponding memento in the original archive: we compare the Memento-Datetimes, the URI-Rs, and the final HTTP status codes, and if any of these values do not match, we declare that it is a \emph{missing memento}.

\section{Findings}

Table \ref{tab:archive-changes} shows the original and new archives, if the Memento-Datetimes matched, if the HTTP status codes matched, and if the URI-Rs matched, along with the number of mementos in each category. The number of mementos we consider as missing are in \textbf{bold}.  
We studied a total of 1,981 mementos from these four archives (we only count the NLI mementos once), classified 537 as missing (i.e., different Memento-Datetimes, status codes, or URI-Rs), and were unable to rediscover any version of 20 mementos in their corresponding new archives (these have NO in all columns in the table).

\begin{table}
\caption{Web archive changes based on how mementos changed. The number of missing mementos is shown in \textbf{bold}.}
\label{tab:archive-changes}
\begin{tabular}{l|ccc|r}
\hline
\textbf{\begin{tabular}[c]{@{}l@{}}Original archive \\ → New archive\end{tabular}}               & \textbf{\begin{tabular}[c]{@{}c@{}}Same \\ {\scriptsize Memento-Datetimes?}\end{tabular}} & \textbf{\begin{tabular}[c]{@{}c@{}}Same status\\ codes?\end{tabular}} & \textbf{\begin{tabular}[c]{@{}c@{}}Same \\ URI-Rs?\end{tabular}} & \textbf{URI-Ms} \\ \hline
\multirow{5}{*}{\begin{tabular}[c]{@{}l@{}}collectionscanada.gc.ca \\ → bac-lac.gc.ca\end{tabular}} 
     & Yes     & Yes    & Yes     & 302              \\
     & NO      & Yes    & Yes      & \textbf{28}     \\
     & NO      & Yes     & NO      & \textbf{18}     \\
     & NO      & NO    & Yes       & \textbf{1}      \\ 
     & NO      & NO     & NO       & \textbf{2}      \\
     \hline
\begin{tabular}[c]{@{}l@{}}europarchive.org/NLI \\ → internetmemory.org/NLI\end{tabular}        & Yes     & Yes    & Yes    & 979             \\ \hline
\multirow{5}{*}{\begin{tabular}[c]{@{}l@{}}internetmemory.org/NLI \\ → archive-it.org\end{tabular}} 
     & Yes      & Yes      & Yes    & 787             \\
     & Yes      & NO      & Yes     & \textbf{1}      \\ 
     & Yes      & NO       & NO     & \textbf{2}      \\
     & NO       & Yes      & Yes    & \textbf{184}    \\
     & NO       & Yes       & NO    & \textbf{5}      \\
     \hline
\multirow{4}{*}{\begin{tabular}[c]{@{}l@{}}proni.gov.uk \\ → archive-it.org\end{tabular}}       & Yes      & Yes     & Yes      & 355             \\
     & Yes      & NO     & Yes       & \textbf{2}      \\ 
     & NO       & Yes     & Yes      & \textbf{106}    \\
     & NO       & Yes      & NO      & \textbf{6}      \\
     \hline
\multirow{2}{*}{\begin{tabular}[c]{@{}l@{}}perma-archives.org \\ → perma.cc\end{tabular}}       & NO       & Yes     & Yes      & \textbf{164}    \\
     & NO       & NO      & NO       & \textbf{18}     \\ 
     \hline
\end{tabular}
\end{table}

\subsection{Library and Archives Canada}

In our study, we had 351 mementos from \url{collectionscanada.gc.ca}, maintained by Library and Archives Canada (LAC). In July 2018 we discovered that all 351 URI-Ms from this archive were redirecting to \url{http://www.bac-lac.gc.ca/eng/discover/archives-web-government/Pages/web-archives.aspx}, the main webpage of the Government of Canada Web Archive. By viewing that live webpage, we discovered that the contents of this web archive had moved to \url{webarchive.bac-lac.gc.ca}. 
Additional details of our findings regarding LAC can be found in our blog post  \cite{aturban-blog-lac}. 

\begin{figure}[ht!]
\centering
\begin{subfigure}{0.42\textwidth}
  \centering
  \fbox{
  	\includegraphics[width=1\linewidth]{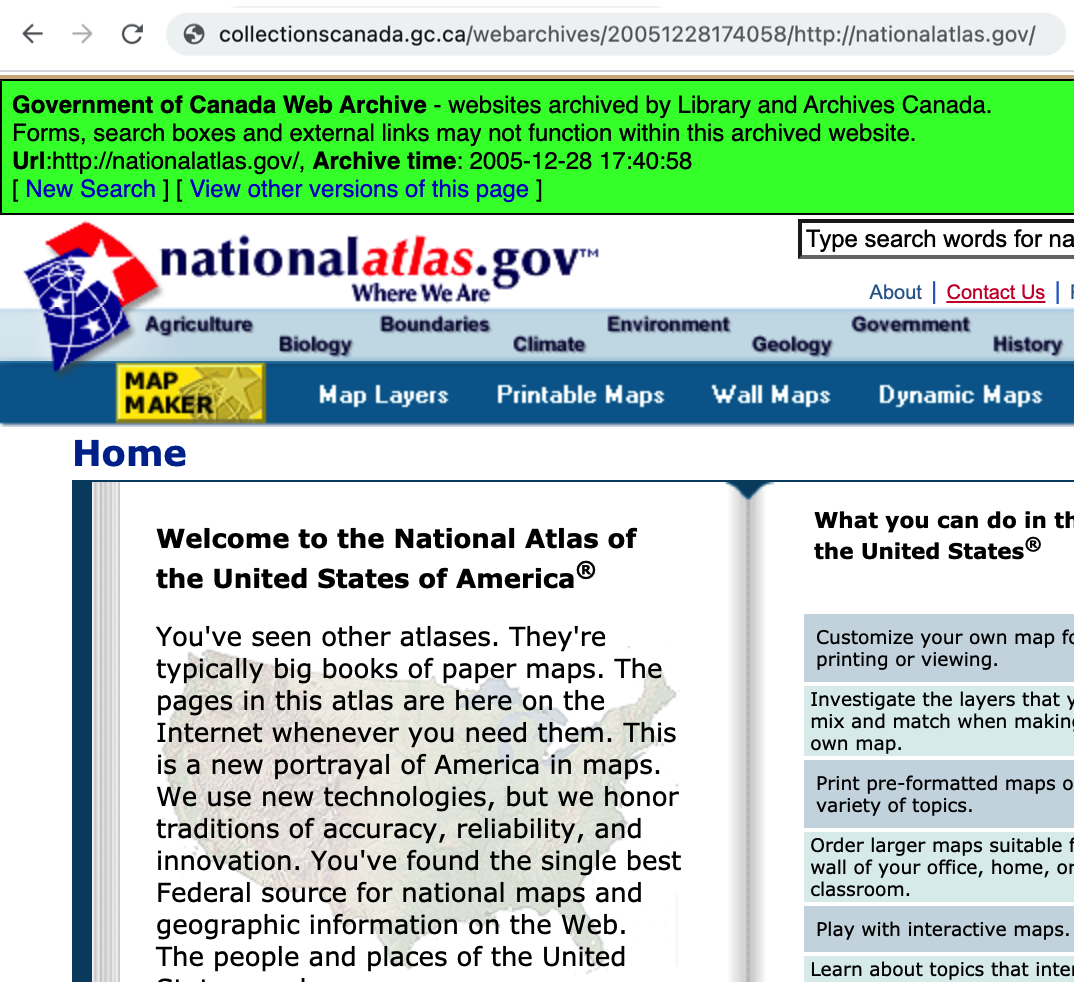}
  }
  \caption{In \url{collectionscanada.gc.ca}}
  \label{fig:lac-1}
\end{subfigure}%
\hspace{25pt}
\begin{subfigure}{0.42\textwidth}
  \centering
  \fbox{
  	\includegraphics[width=1\linewidth]{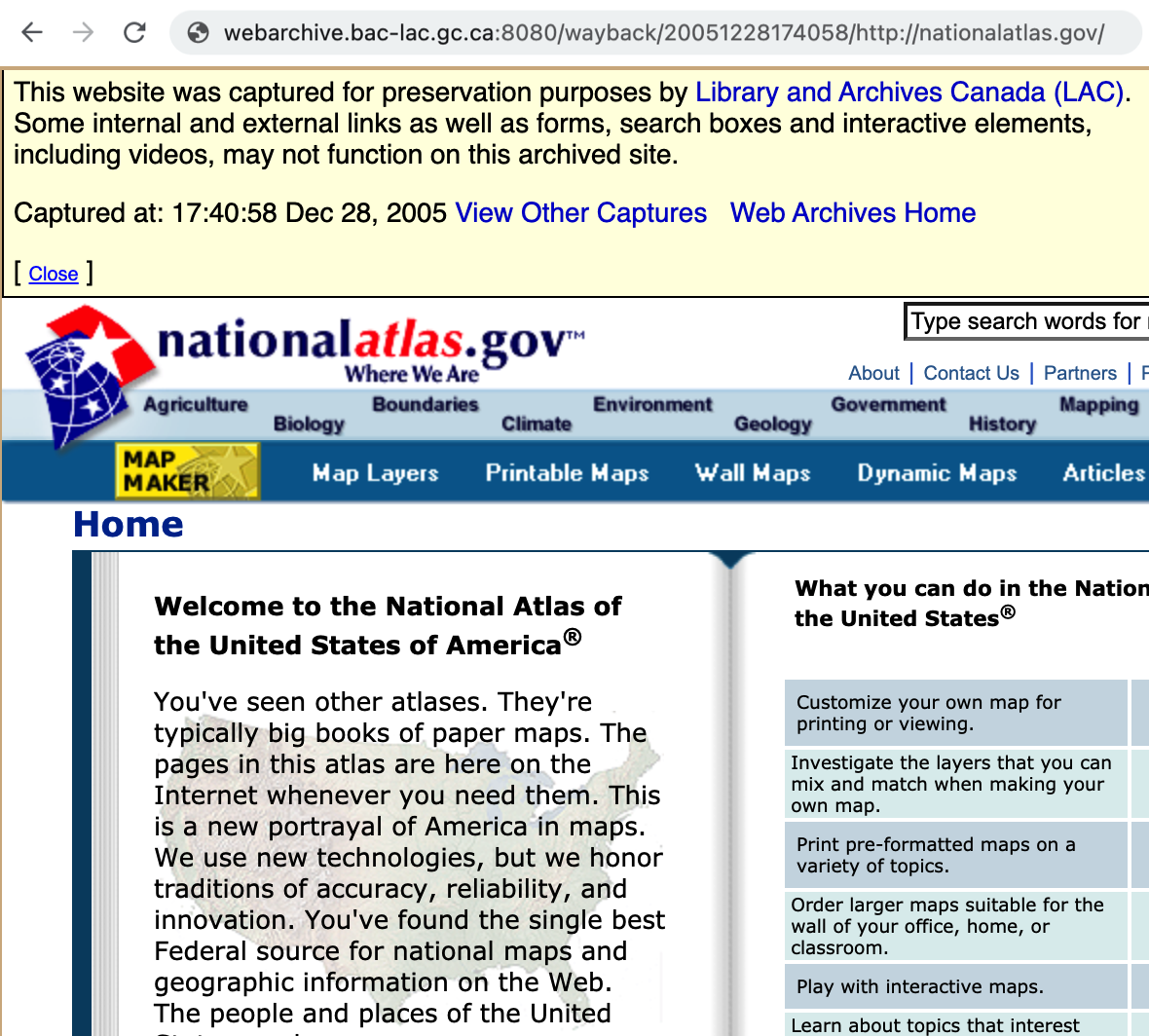}
  }
  \caption{In \url{webarchive.bac-lac.gc.ca}}
  \label{fig:lac-2}
\end{subfigure}
\caption{An example of a memento moved from \url{collectionscanada.gc.ca} to \url{webarchive.bac-lac.gc.ca}.}
\label{fig:lac}
\end{figure}


Because LAC still controls the domain of the original archive
(\url{collectionscanada.gc.ca}), it would be possible for requests to the original archive to redirect to the corresponding URI-Ms in the new archive, for example, using the Apache \texttt{mod\_rewrite} rules shown in Figure \ref{fig:rules-ca}.

\begin{figure}[ht!]
\begin{lstlisting}
# With mod_rewrite
RewriteEngine on
RewriteRule   "^/webarchives/(\d{14})/(.+)" 
              http://webarchive.bac-lac.gc.ca:8080/wayback/$1/$2  
              [L,R=301]
\end{lstlisting}
\caption{The Apache \texttt{mod\_rewrite} rules that can be used to handle redirects from \url{www.collectionscanada.gc.ca} to \url{webarchive.bac-lac.gc.ca}.} 
\label{fig:rules-ca}
\end{figure}

 This would maintain link integrity via ``follow-your-nose" \cite{fielding:apis} from the old URI-M to the new URI-M. 



But since we found that every memento request to the original archive redirected to the home page of the new archive, we had to manually intervene to detect the corresponding URI-Ms of the mementos in the new archive.  This was done by replacing \url{www.collectionscanada.gc.ca/webarchives} with \url{webarchive.bac-lac.gc.ca:8080/wayback} in the URI-Ms of the original archive. For instance,  \url{http://www.collectionscanada.gc.ca/webarchives/20051228174058/http://nationalatlas.gov/} 
is now available at
\url{http://webarchive.bac-lac.gc.ca:8080/wayback/20051228174058/http://nationalatlas.gov/}


Many of the mementos from LAC have been archived by the Internet Archive, meaning that the URI-Ms are archived, not just the URI-Rs.  For instance, \url{http://web.archive.org/web/20160720232234/http://www.collectionscanada.gc.ca/webarchives/20071125005256/http://www.phac-aspc.gc.ca/publicat/ccdr-rmtc/95vol21/index.html} is a URI-M captured in the Internet Archive in July 2016 of a URI-M captured by LAC in November 2007.  Because of this, we were able to estimate when LAC made the change from \url{www.collectionscanada.gc.ca} to \url{webarchive.bac-lac.gc.ca}.  It appears that LAC began using the new archive around Dec 2011, with \url{http://web.archive.org/web/20111211144417/http://www.collectionscanada.gc.ca/} linking to \url{http://web.archive.org/web/20111207015200/http://www.bac-lac.gc.ca/eng/Pages/default.aspx}. Starting in February 2017, the Internet Archive was capturing mementos from both \url{webarchive.bac-lac.gc.ca:8080/wayback/} and  \url{collectionscanada.gc.ca/webarchives/}, indicating that LAC had two separate archives operational concurrently, but only URI-Ms from \url{collectionscanada.gc.ca} were being returned to the Memento LANL Aggregator.

%
%













\begin{figure}[ht!]
\centering
\fbox{
\includegraphics[width=0.90\textwidth, trim =0 40 40 50, clip]{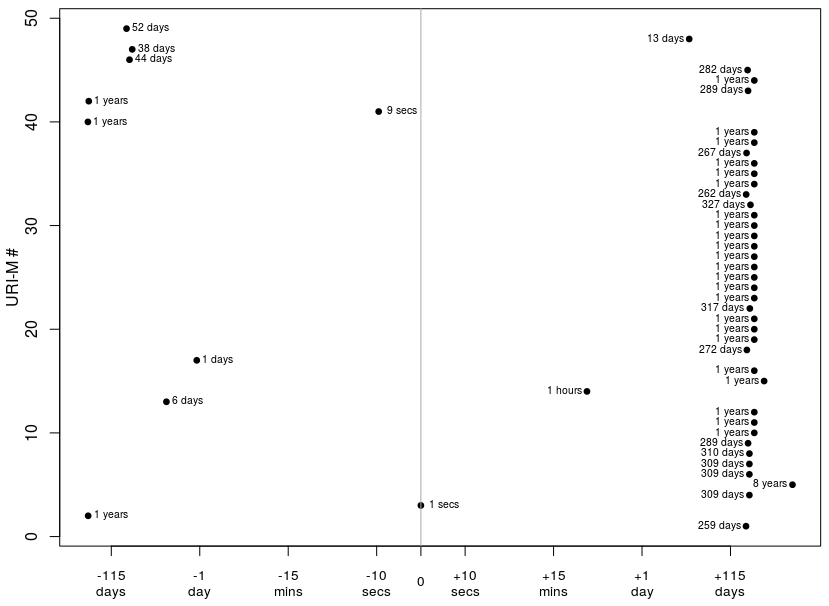}
}
\caption{Difference between the Memento-Datetimes for URI-Ms from \url{www.collectionscanada.gc.ca} and the corresponding URI-Ms from \url{webarchive.bac-lac.gc.ca}.} 
\label{fig:diff-ca}
\end{figure}

We classified 49 out of 351 mementos from \url{www.collectionscanada.gc.ca} as missing because they cannot be retrieved exactly from the new archive as they were in the old archive. Instead, the new archive responds with other mementos that have different Memento-Datetimes. 
For example, when we requested the URI-M
\url{http://www.collectionscanada.gc.ca/webarchives/20060208075019/http://www.cdc.gov/}
from the original archive on February 27, 2018, we received the HTTP status \texttt{200 OK} and a Memento-Datetime of Wed, 08 Feb 2006 07:50:19 GMT.  Then, we requested the corresponding URI-M, 
\url{http://webarchive.bac-lac.gc.ca:8080/wayback/20060208075019/http://www.cdc.gov/},
from the new archive. This request is redirected to another URI-M, 
\url{http://webarchive.bac-lac.gc.ca:8080/wayback/20061026060247/http://www.cdc.gov/}, which has a different Memento-Datetime (Thu, 26 Oct 2006 06:02:47 GMT), resulting in a delta of about 260 days. In addition, the content of the memento in the new archive is different from the content of the memento from the original archive. Figure \ref{fig:diff-ca} shows the difference between the original Memento-Datetime and the new Memento-Datetime for each URI-M. 

We also found that the HTTP status codes of URI-Ms in the new archive might not be identical to the HTTP status code of the corresponding URI-Ms in the original archive. For example, the HTTP request for the URI-M \url{http://www.collectionscanada.gc.ca/webarchives/20070220181041/http://www.berlin.gc.ca/} from the original archive resulted in the multiple HTTP \texttt{302} redirects before ending up with the HTTP status code \texttt{404} (Figure \ref{fig:redirects-404}).

\begin{figure}[ht!]
\begin{lstlisting}
http://www.collectionscanada.gc.ca/webarchives/20070220181041/http://www.berlin.gc.ca/ (302)
http://www.collectionscanada.gc.ca/webarchives/20070220181041/http://www.dfait-maeci.gc.ca/canadaeuropa/germany/ (302)
http://www.collectionscanada.gc.ca/webarchives/20070220181204/http://www.dfait-maeci.gc.ca/canadaeuropa/germany/ (302)
http://www.collectionscanada.gc.ca/webarchives/20070220181204/http://www.international.gc.ca/global/errors/404.asp?404%3Bhttp://www.dfait-maeci.gc.ca/canadaeuropa/germany/ (404)
\end{lstlisting}
\caption{The HTTP status codes of the URI-M \url{http://www.collectionscanada.gc.ca/webarchives/20070220181041/http://www.berlin.gc.ca/} from the original archive.} 
\label{fig:redirects-404}
\end{figure}

When we requested the corresponding URI-M from the new archive, it ended up with the HTTP status code \texttt{200} (Figure \ref{fig:redirects-200}):

\begin{figure}[ht!]
\begin{lstlisting}
http://webarchive.bac-lac.gc.ca:8080/wayback/20070220181041/http://www.berlin.gc.ca/ (Redirect by JavaScript (JS))
http://webarchive.bac-lac.gc.ca:8080/wayback/20070220181041/http://www.dfait-maeci.gc.ca/canadaeuropa/germany/ (Redirect by JS)
http://webarchive.bac-lac.gc.ca:8080/wayback/20070220181204/http://www.international.gc.ca/global/errors/404.asp?404%3Bhttp://www.dfait-maeci.gc.ca/canadaeuropa/germany/ (Redirect by JS)
http://webarchive.bac-lac.gc.ca:8080/wayback/20071115025620/http://www.international.gc.ca/canada-europa/germany/ (302) 
http://webarchive.bac-lac.gc.ca:8080/wayback/20071115023828/http://www.international.gc.ca/canada-europa/germany/ (200) 
\end{lstlisting}
\caption{The HTTP status codes of the URI-M \url{http://webarchive.bac-lac.gc.ca:8080/wayback/20070220181041/http://www.berlin.gc.ca/} from the new archive.} 
\label{fig:redirects-200}
\end{figure}


Additionally, as of this writing, it appears that all URI-Ms in the new archive are redirecting to \url{https://www.bac-lac.gc.ca/eng/discover/archives-web-government/Pages/web-archives.aspx} which states that ``the Government of Canada Web Archive is currently not available".  This message has been displayed on the webpage since April 2020\footnote{Although this is outside of our 14-month study, this effectively means that all 351 LAC mementos are currently missing.}. 

\subsection{National Library of Ireland}

\begin{figure}[ht!]
\centering
\begin{subfigure}{0.5\textwidth}
  \centering
  \fbox{
  	\includegraphics[width=160pt, height=150pt]{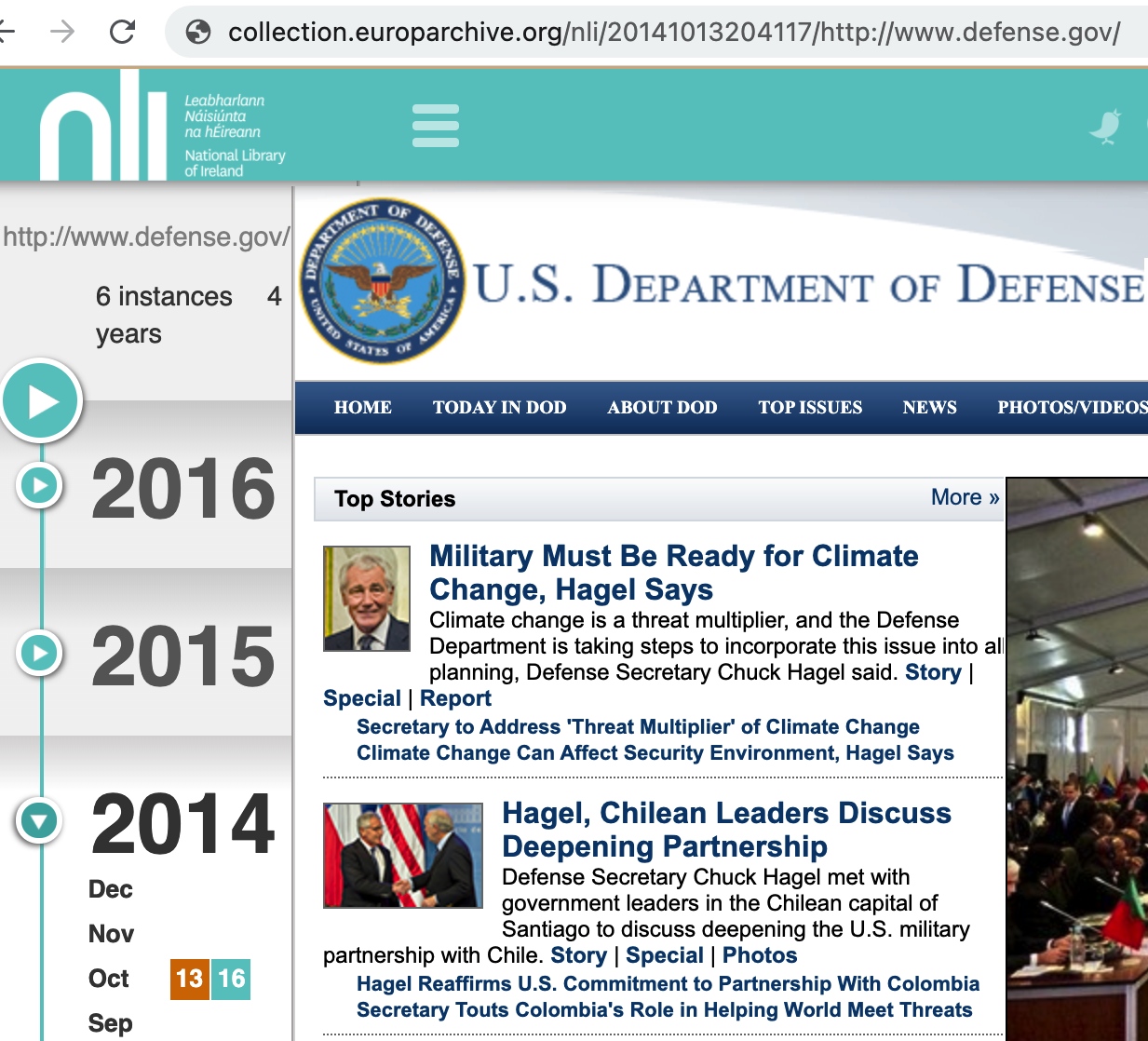}
  }
  \caption{In \url{europarchive.org}\\\hspace{\textwidth}}
  \label{fig:nli-1}
\end{subfigure}%
\begin{subfigure}{0.5\textwidth}
  \centering
  \fbox{
  	\includegraphics[width=160pt, height=150pt]{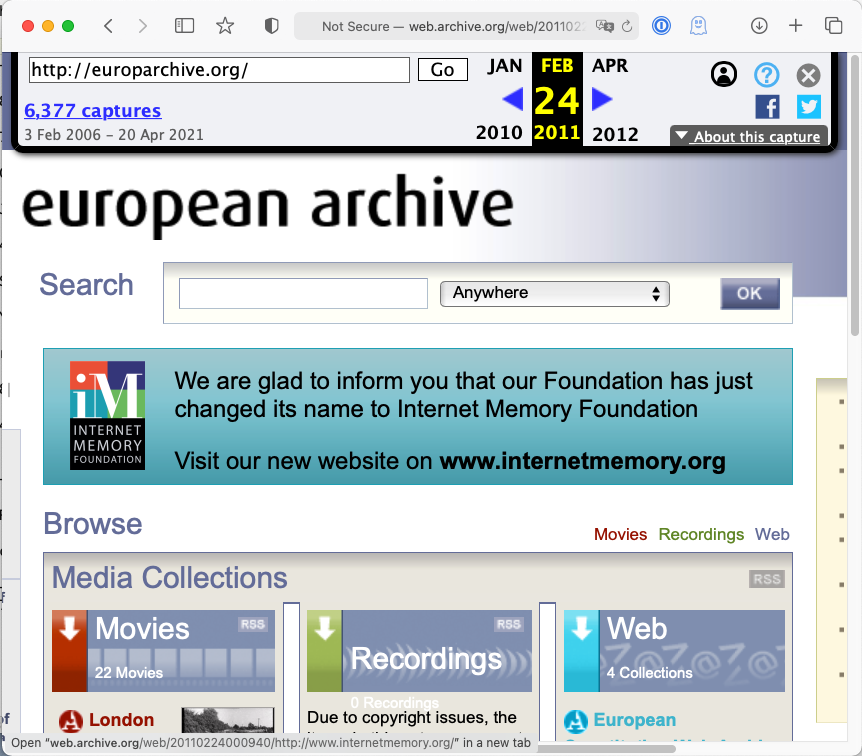}
  }
  \caption{European Archive announcing their name change to IMF}
  \label{fig:europarchive-imf}
\end{subfigure}

\begin{subfigure}{0.5\textwidth}
  \centering
  \fbox{
  	\includegraphics[width=160pt, height=150pt]{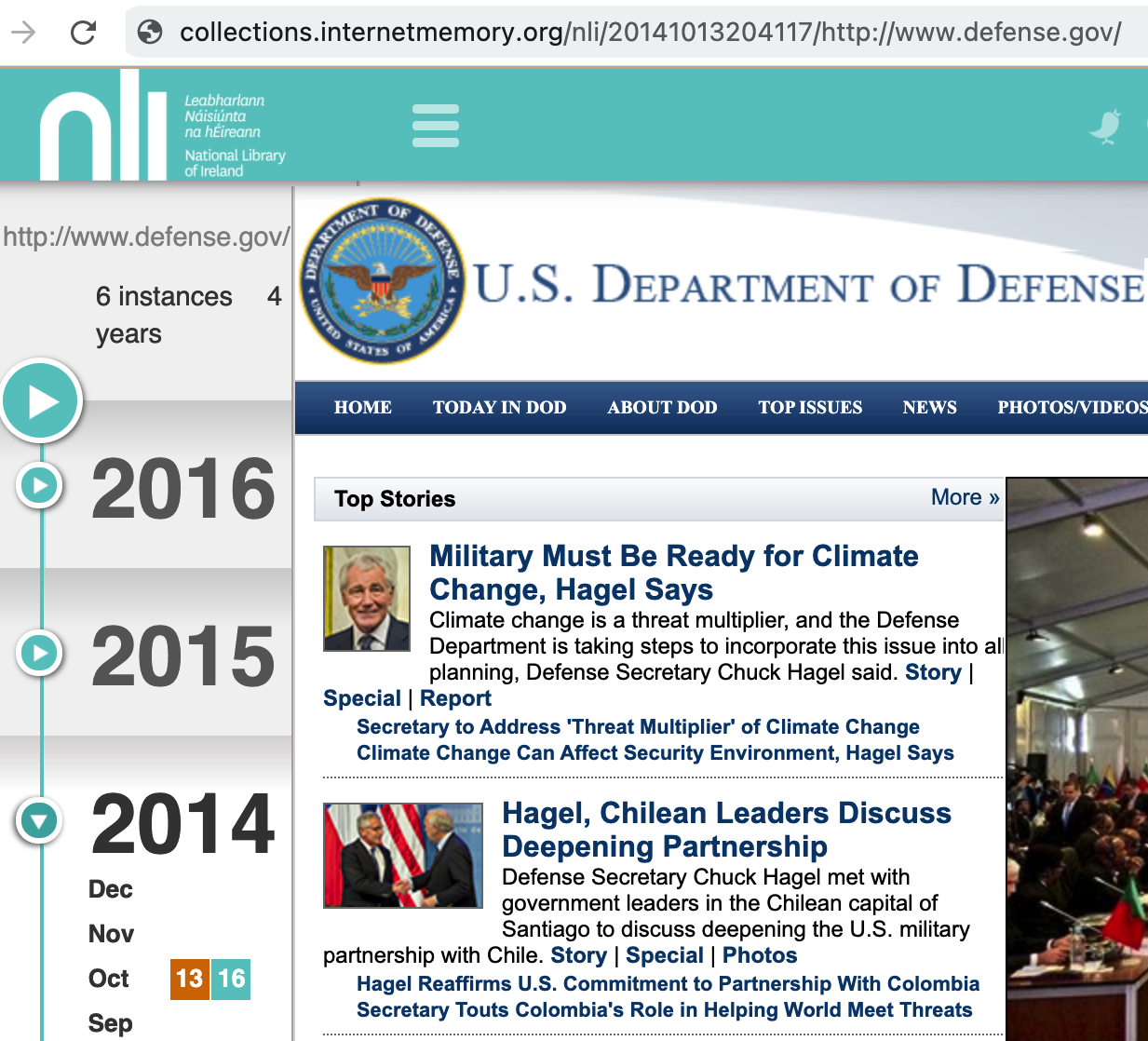}
  }
  \caption{In \url{internetmemory.org}}
  \label{fig:nli-2}
\end{subfigure}%
\begin{subfigure}{0.5\textwidth}
  \centering
  \fbox{
  	\includegraphics[width=160pt, height=150pt]{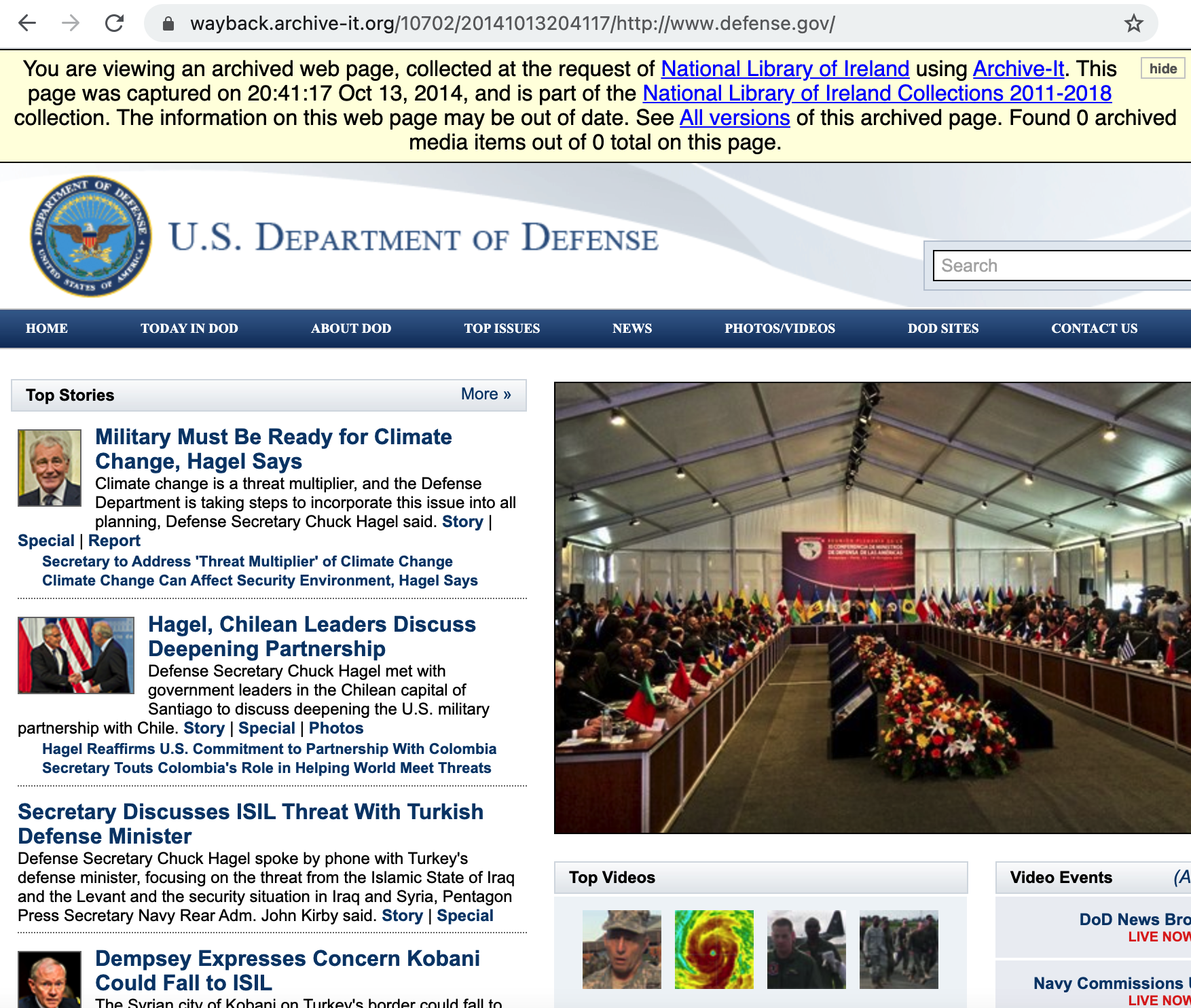}
  }
  \caption{In \url{archive-it.org} }
  \label{fig:nli-3}
\end{subfigure}%

\caption{An example of a memento moved from \url{europarchive.org} to \url{internetmemory.org}, and then to \url{archive-it.org}}
\label{fig:nli}
\end{figure}

In May 2018, we discovered that 979 mementos from the National Library of Ireland (NLI) collection that were originally hosted by The European Archive at \url{europarchive.org} were moved to \url{internetmemory.org}, hosted by the Internet Memory Foundation. This appears to have just been a domain name change, as The European Archive announced its name change to the Internet Memory Foundation in 2011 (Figure \ref{fig:europarchive-imf}), but had been still returning URI-Ms with the domain \url{europarchive.org} to the LANL Memento Aggregator. Although there was a human-readable notice that the domain name would be changing, there was no machine-readable notice provided. In addition to using \texttt{mod\_rewrite} to provide automatic redirects, another option would be to use the Sunset HTTP response header \cite{sunset:rfc} on requests for URI-Ms from the original archive. In September 2018, we found that the collection of mementos had been moved to Archive-It (\url{archive-it.org}) in collection \url{https://archive-it.org/collections/10702}.  Figure \ref{fig:nli} shows a single memento represented in the three different archives. 
Additional details of our findings regarding NLI can be found in our blog post  \cite{aturban-blog-nli}. 

There were no changes in the 979 mementos (other than their URIs) when they moved from \url{europarchive.org} to \url{internetmemory.org}, thus our assumption that this was only a domain name change. As with LAC, the European Archive did not use URL rewriting to automatically redirect requests for URI-Ms with the \url{collection.europarchive.org} domain to \url{collections.internetmemory.org}; we had to manually make the changes in our dataset. In addition after the changeover, the main webpage for \url{europarchive.org} itself was no longer maintained, and by August 2018 it had been taken over by an apparent spam site.  Currently, the main website for \url{internetmemory.org} does not respond, and it has not been archived by the Internet Archive's Wayback Machine since January 2019.  



The movement of mementos from these archives will affect link integrity across web resources that contain links to mementos from \url{europarchive.org} and \url{internetmemory.org}. In addition, as shown in Figure \ref{fig:nli}, the custom archive banners of the European Archive and IMF appear to be the same. Via these banners, the archives allow users to view the number of available mementos per year and the representation of a selected memento in the same page. Archive-It, on the other hand, uses a basic playback banner as shown in Figure \ref{fig:nli}. 

We found that upon moving to Archive-It, 192 of the original 979 mementos were missing and cannot be retrieved from the new archive. 
For these missing mementos, the new archive responds with other mementos that have different values for the Memento-Datetime, the URI-R, or the HTTP status code. One example shows a memento that cannot be found in the new archive  with the same Memento-Datetime as it was in the original archive. When requesting the URI-M
\url{http://collections.internetmemory.org/nli/20121221162201/http://bbc.co.uk/news/}
from the original archive on September 3, 2018, the archive responded with \texttt{200 OK}, and the Memento-Datetime was Fri, 21 Dec 2012 16:22:01 GMT. Then, we requested the corresponding URI-M,
\url{http://wayback.archive-it.org/10702/20121221162201/http://bbc.co.uk/news/},
from the new archive. The request redirected to another URI-M \url{http://wayback.archive-it.org/10702/20121221163248/http://www.bbc.co.uk/news/}, as shown in Figure \ref{fig:redirects-it-bbc-200-ok}.

\begin{figure}[ht!]
\begin{lstlisting}
$ curl --head --location --silent http://wayback.archive-it.org/10702/20121221162201/http://bbc.co.uk/news/ | egrep -i "(HTTP/|^location:|^Memento-Datetime)"

HTTP/1.1 302 Found
Location: /10702/20121221163248/http://www.bbc.co.uk/news/
HTTP/1.1 200 OK
Memento-Datetime: Fri, 21 Dec 2012 16:32:48 GMT
\end{lstlisting}
\caption{The HTTP status codes of the URI-M \url{http://wayback.archive-it.org/10702/20121221162201/http://bbc.co.uk/news/} from the new archive \url{archive-it.org}.} 
\label{fig:redirects-it-bbc-200-ok}
\end{figure}

\begin{figure}[ht!]
\centering
\fbox{
\includegraphics[width=0.90\textwidth]{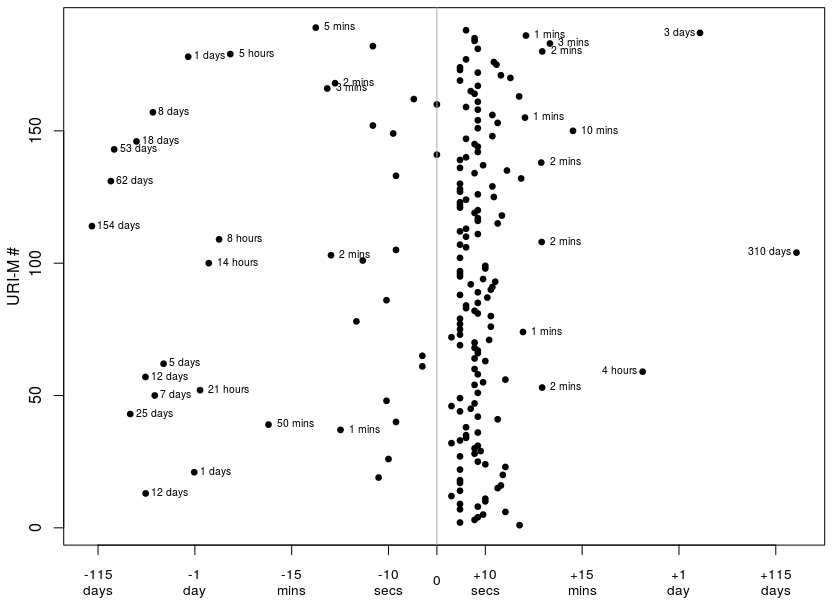}
}
\caption{Difference between the Memento-Datetimes for URI-Ms from \url{internetmemory.org} and the corresponding URI-Ms from \url{archive-it.org}.} 
\label{fig:diff-nli}
\end{figure}

Although the representations of both mementos are identical (except for the archival banners), we consider the memento from the original archive as missing because both mementos have different values for Memento-Datetime (i.e., Fri, 21 Dec 2012 16:32:48 GMT in the new archive) for a delta of about 10 minutes. Even though the 10-minute delta might not be semantically significant (apparently just a change in the canonicalization of the URI-R, with \url{bbc.co.uk} redirecting to \url{www.bbc.co.uk}), we do not consider it to be the same since the values of the Memento-Datetime are not identical. Figure \ref{fig:diff-nli} shows the difference between the original Memento-Datetime and the new Memento-Datetime for each URI-M. 

The second example shows a memento that has different values for the Memento-Datetime and URI-R compared to the corresponding values from the original archive. When requesting the memento \url{http://collections.internetmemory.org/nli/20121223122758/http://www.whitehouse.gov/} on September 03, 2018, the original archive returned \texttt{200 OK} for an archived \texttt{403 Forbidden}. When requesting the corresponding memento from archive-it.org  \url{http://wayback.archive-it.org/10702/20121223122758/http://www.whitehouse.gov/}, the request redirected to another URI-M \url{http://wayback.archive-it.org/10702/20121221222130/http://www.whitehouse.gov/administration/eop/nec/speeches/gene-sperling-remarks-economic-club-washington}, which is \texttt{200 OK}. Not only are the values of the Memento-Datetime different, but also the URI-Rs. 

%


Another change we found was in the way the archives handled archived HTTP \texttt{4xx/5xx} status codes. The replay tool in the original archive  was configured so that it returned the status code \texttt{200 OK} for archived \texttt{4xx/5xx}. 
For example, when requesting the memento \url{http://collections.internetmemory.org/nli/20121021203647/http://www.amazon.com/} on September 03, 2018, the original archive returned \texttt{200 OK} for an archived \texttt{503 Service Unavailable}.  Even the HTTP status code of the inner iframe in which the archived content is loaded returned \texttt{200 OK}.  When requesting the corresponding memento \url{http://wayback.archive-it.org/10702/20121021203647/http://www.amazon.com/}, 
as described in Section \ref{sec:background}, Archive-It properly returns the status code \texttt{503 Service Unavailable} for an archived \texttt{503} response.

\begin{figure}[ht!]
\centering
\fbox{
\includegraphics[width=0.9\textwidth]{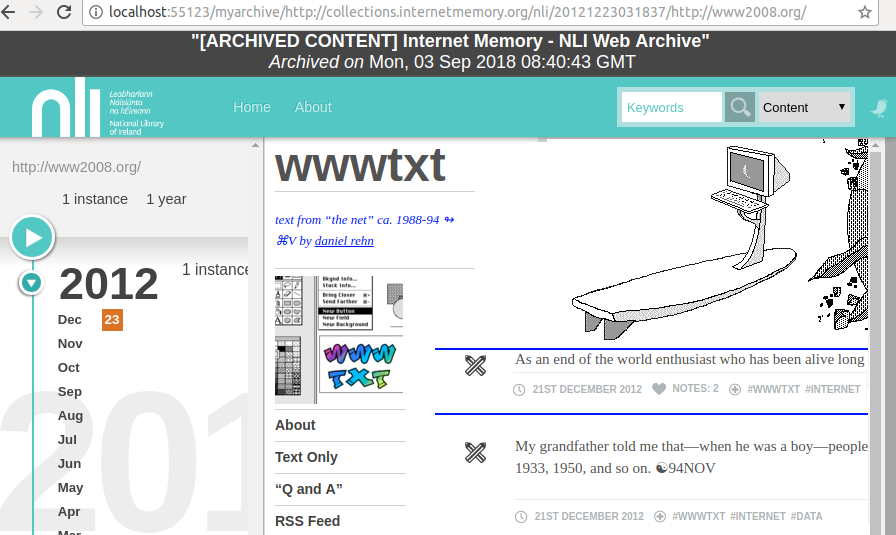}
}

\caption{The representation of a memento (URI-M = \url{http://collections.internetmemory.org/nli/20121223031837/http://www2008.org/}) from the IMF.} 
\label{fig:www2008-imf}
\end{figure}

Finally, we found that some HTTP status codes of URI-Ms in the new archive might not be identical to the HTTP status code of the corresponding URI-Ms in the original archive. For example, the HTTP request of the URI-M \url{http://collections.internetmemory.org/nli/20121223031837/http://www2008.org/} to the original archive resulted in \texttt{200 OK} (Figure \ref{fig:www2008-imf}).  The request to the corresponding URI-M \url{http://wayback.archive-it.org/10702/20121223031837/http://www2008.org/} from Archive-It results in \texttt{404 Not Found} as the cURL session in Figure \ref{fig:redirects-it-www2008-404} shows:

\begin{figure}[ht!]
\begin{lstlisting}
$ curl --head --silent http://wayback.archive-it.org/10702/20121223031837/http://www2008.org/

HTTP/1.1 404 Not Found
Server: Apache-Coyote/1.1
Content-Security-Policy-Report-Only: default-src "self"
[...]
Content-Type: text/html;charset=utf-8
Content-Length: 4902
Date: Thu, 05 Sep 2019 08:28:27 GMT
\end{lstlisting}
\caption{The HTTP status codes of the URI-M \url{http://wayback.archive-it.org/10702/20121223031837/http://www2008.org/} from the new archive \url{archive-it.org}.} 
\label{fig:redirects-it-www2008-404}
\end{figure}

We note that the move to Archive-It included a change of web archiving platform, which can affect the replay of mementos. Therefore, when the original WARC files were moved to the new platform, differences in indexing, replay, URI canonicalization, and handling of HTTP redirections  may explain some of the differences in the values of Memento-Datetime, URI-R, and HTTP status code in the new archive.

\subsection{Public Record Office of Northern Ireland (PRONI)}

\begin{figure}[ht!]
\centering
\fbox{
\includegraphics[width=0.9\textwidth]{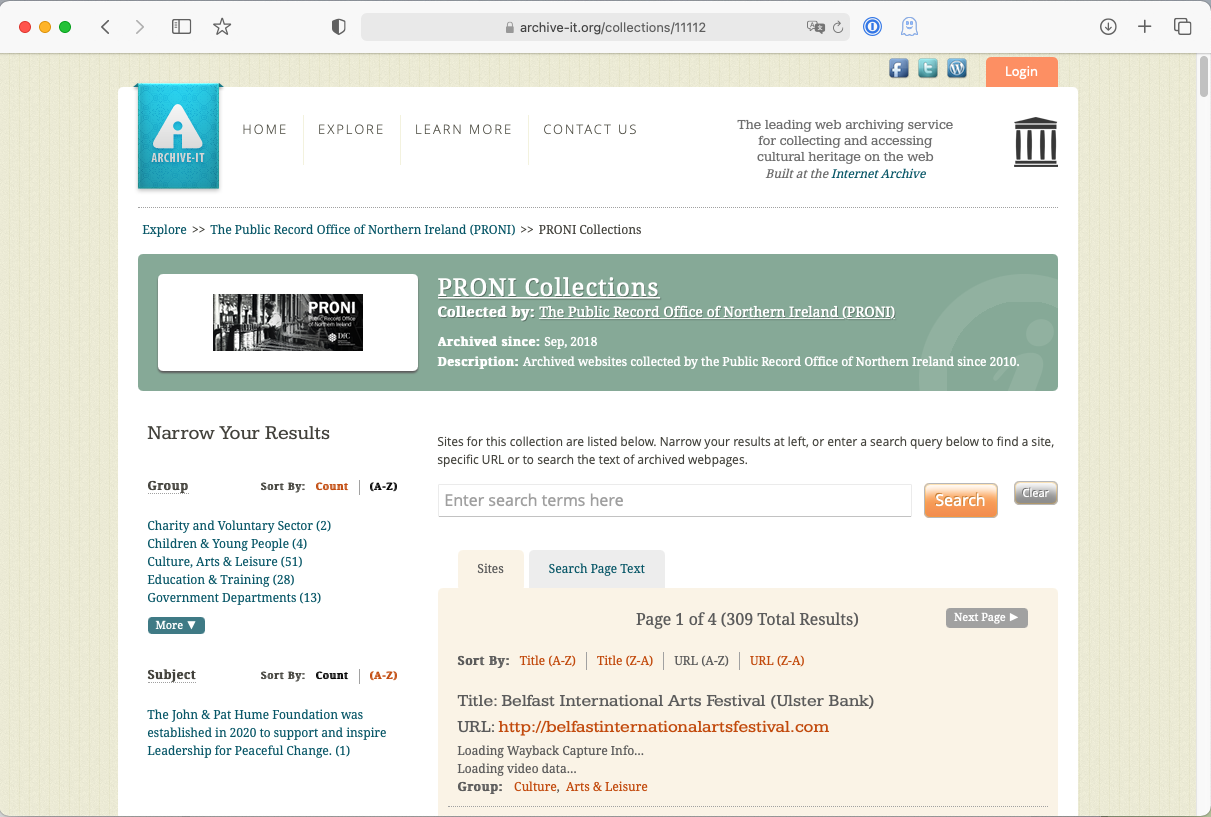}
}
\caption{The homepage of the PRONI collection at Archive-It.} 
\label{fig:proni-archiveit}
\end{figure}

The Public Record Office of Northern Ireland (PRONI) Web Archive was also hosted by the European Archive/IMF, but using a custom domain, \url{webarchive.proni.gov.uk}. In October 2018, mementos in the PRONI archive were moved to Archive-It (\url{archive-it.org}) in the collection at \url{https://archive-it.org/collections/11112} as shown in Figure \ref{fig:proni-archiveit}. 
After the move, we found that 114 of the original 469 mementos in our study were missing. Additional details of our findings regarding PRONI can be found in our blog post \cite{aturban-blog-proni}. 

As with LAC and NLI, requests for URI-Ms in PRONI do not automatically redirect to mementos at Archive-It. In fact, every request to a URI-M in the PRONI archive now returns a \texttt{404 Not Found} status code, as the cURL session in Figure \ref{fig:every-request-to-proni} shows. 

\begin{figure}[ht!]
\begin{lstlisting}
$ curl --head --location http://webarchive.proni.gov.uk/20100218151844/http://www.berr.gov.uk/

HTTP/1.1 302 Found
Cache-Control: no-cache
Content-length: 0
Location: https://webarchive.proni.gov.uk/20100218151844/http://www.berr.gov.uk/
HTTP/2 404
date: Fri, 20 Sep 2019 08:13:45 GMT
server: Apache/2.4.18 (Ubuntu)
content-type: text/html; charset=iso-8859-1
\end{lstlisting}
\caption{The HTTP status codes of the URI-M \url{http://webarchive.proni.gov.uk/20100218151844/http://www.berr.gov.uk/} from the original archive \url{webarchive.proni.gov.uk}.} 
\label{fig:every-request-to-proni}
\end{figure}

\begin{figure}[ht!]
\centering
\begin{subfigure}{0.45\textwidth}
  \centering
  \fbox{
  	\includegraphics[width=150pt, height=140pt]{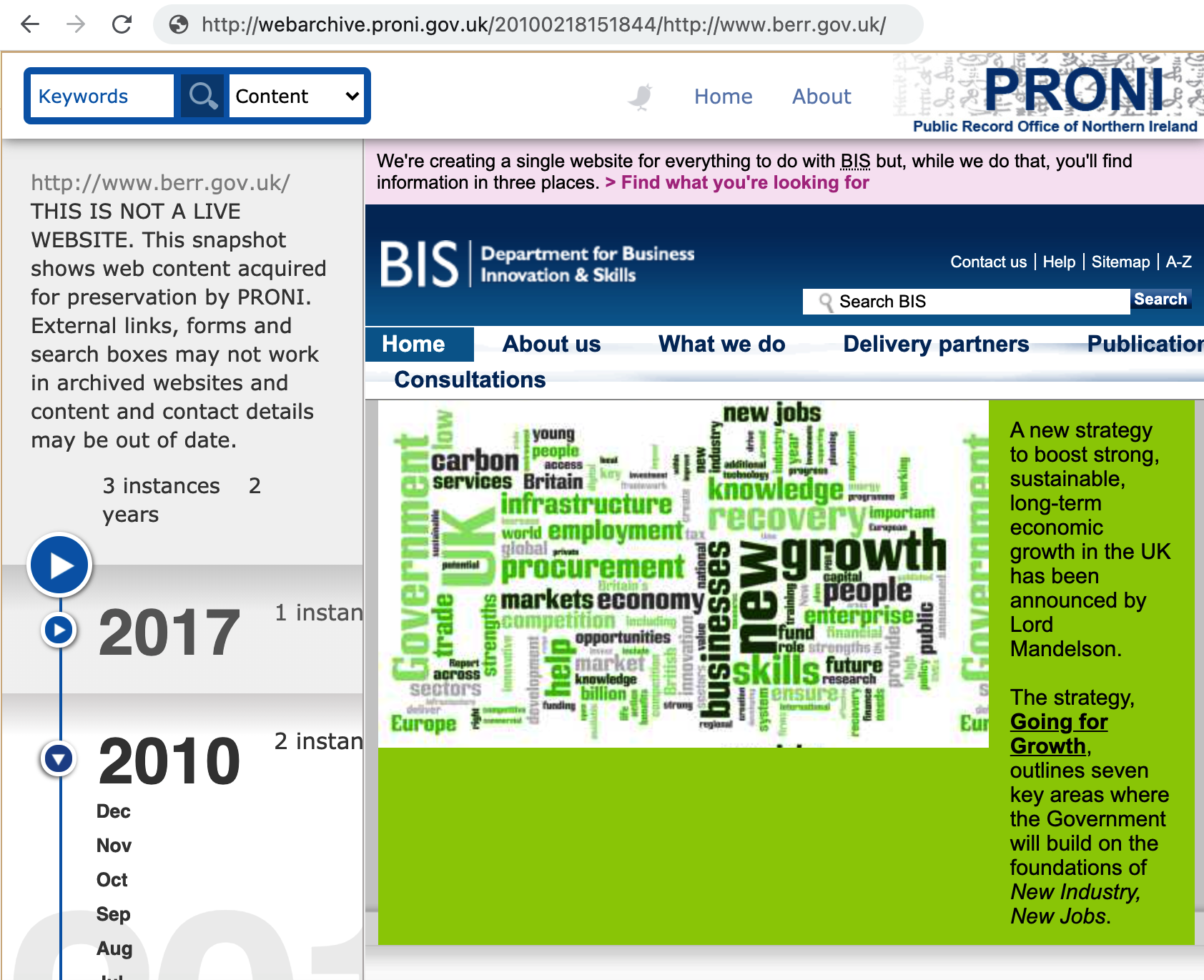}
  }
  \caption{In \url{webarchive.proni.gov.uk}}
  \label{fig:proni-1}
\end{subfigure}%
\hspace{12pt}
\begin{subfigure}{0.45\textwidth}
  \centering
  \fbox{
  	\includegraphics[width=150pt, height=140pt]{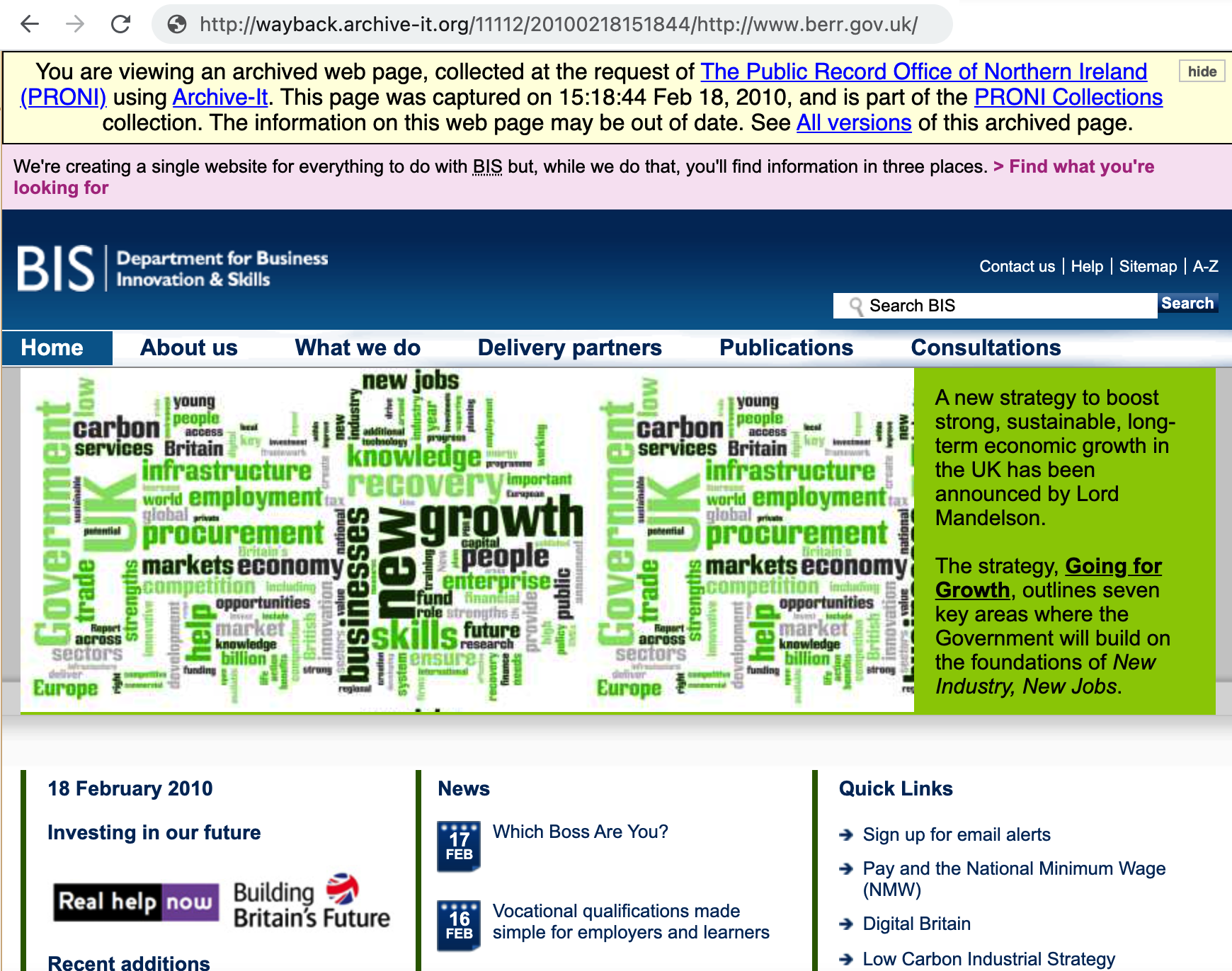}
  }
  \caption{In \url{archive-it}}
  \label{fig:proni-2}
\end{subfigure}
\caption{An example of a memento moved from \url{webarchive.proni.gov.uk} to \url{archive-it.org}.}
\label{fig:proni}
\end{figure}

However, users of the archive can indirectly find the corresponding URI-Ms because \url{https://webarchive.proni.gov.uk} provides a list of the URI-Rs for which mementos have been created.  Let us consider finding the corresponding memento in Archive-It for the PRONI memento
\url{http://webarchive.proni.gov.uk/20150318223351/http://www.afbini.gov.uk/}, 
which has a URI-R of \url{http://www.afbini.gov.uk/} and a Memento-Datetime of Wed 18 Mar 2015 22:33:51 GMT.
From the index at \url{webarchive.proni.gov.uk}, we can click on the URI-R \url{www.afbini.gov.uk}, which will redirect to an Archive-It HTML page that contains all available mementos for the selected URI-R. Then, we choose 2015-03-18, the same Memento-Datetime as in the original archive. 
In addition, once the Archive-It collection ID (11112) is known, URI-Ms from PRONI can be transformed to corresponding Archive-It URI-Ms. For example, the memento
\url{http://webarchive.proni.gov.uk/20100218151844/http://www.berr.gov.uk/}
is now available at
\url{http://wayback.archive-it.org/11112/20100218151844/http://www.berr.gov.uk/}. The representations of both mementos are illustrated in Figure \ref{fig:proni}.

Even though the PRONI collection was hosted by the European Archive, its URI-Ms did not change when \url{europarchive.org} became \url{internetmemory.org}.  It appears that PRONI served mementos under \url{proni.gov.uk} while using the hosting services provided by the European Archive/IMF. Thus, the regular users of the PRONI archive did not notice any change in URI-Ms. We do not believe custom domains are available with Archive-It, so PRONI was unable to continue to host their mementos in their own URI namespace.

Unlike the European Archive and IMF, PRONI still owns the domain name of the original archive, \url{webarchive.proni.gov.uk}. Therefore, to maintain link integrity via ``follow-your-nose", PRONI could issue redirects (even though it currently does not) to the corresponding URI-Ms in Archive-It. For example, since PRONI uses the Apache web server, the \texttt{mod\_rewrite} rule could be used to perform automatic redirects as shown in Figure \ref{fig:rules-proni}.

\begin{figure}[ht!]
\begin{lstlisting}
# With mod_rewrite
RewriteEngine on
RewriteRule "^/(\d{14})/(.+)" 
            http://wayback.archive-it.org/11112/$1/$2 [L,R=301]
\end{lstlisting}
\caption{The Apache \texttt{mod\_rewrite} rules that can be used to handle redirects from \url{webarchive.proni.gov.uk} to \url{archive-it.org}.} 
\label{fig:rules-proni}
\end{figure}

\begin{figure}[ht!]
\centering
\fbox{
\includegraphics[width=0.90\textwidth]{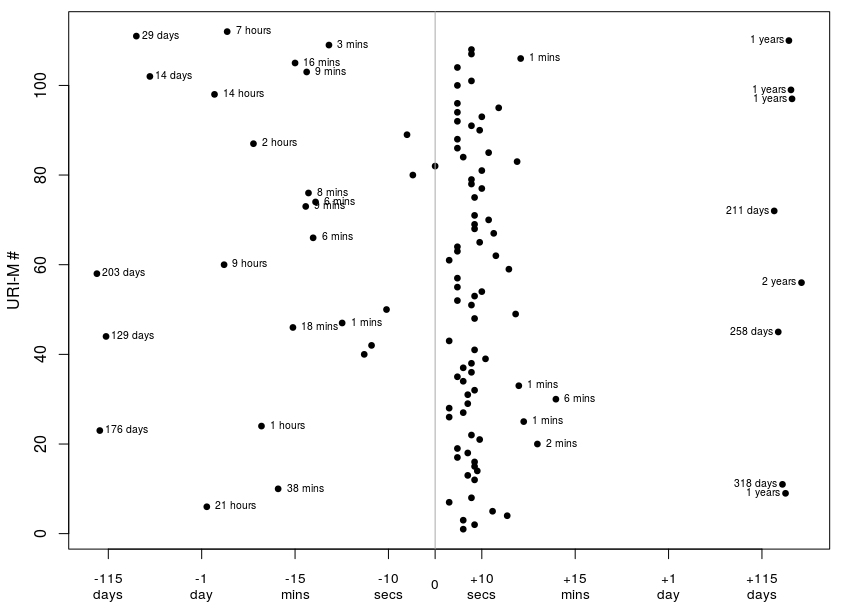}
}
\caption{Difference between the Memento-Datetimes for URI-Ms from \url{webarchive.proni.gov.uk} and the corresponding URI-Ms from \url{archive-it.org}.} 
\label{fig:diff-proni}
\end{figure}

For the 114 missing mementos, the new archive responds with other mementos that have different values for the Memento-Datetime, the URI-R, or the HTTP status code. For example, when requesting the URI-M
\url{http://webarchive.proni.gov.uk/20160901021637/https://www.flickr.com/}
from the original archive on Dec 1, 2017, the archive responded with \texttt{200 OK}, and a Memento-Datetime of Thu, 01 September 2016 02:16:37 GMT. When we requested the corresponding URI-M
\url{http://wayback.archive-it.org/11112/20160901021637/https://www.flickr.com/}
from the new archive, the request was redirected to another URI-M, 
\url{http://wayback.archive-it.org/11112/20170401014520/https://www.flickr.com/}.
The representations of both mementos are identical (except for the archival banners). However, these mementos have different Memento-Datetime values (i.e., Fri, 21 Apr 2017 01:45:20 GMT in the new archive) for a delta of about 211 days. Figure \ref{fig:diff-proni} shows the difference between the original Memento-Datetime and the new Memento-Datetime for each URI-M. 

We found that 63 of the 114 missing mementos have Memento-Datetime values with a delta of less than 11 seconds. For example, the request to the memento
\url{http://webarchive.proni.gov.uk/20170102004044/http://www.fws.gov/}
from the original archive on November 18, 2017 returned a \texttt{302} redirect to
\url{http://webarchive.proni.gov.uk/20170102004044/https://fws.gov/}. 
The request to the corresponding memento
\url{http://wayback.archive-it.org/11112/20170102004044/http://www.fws.gov/}
from the new archive redirects to the memento
\url{http://wayback.archive-it.org/11112/20170102004051/https://www.fws.gov/}.
There is a 10-second difference between the values of the Memento-Datetimes, which might not be semantically significant, 
but we do not consider the mementos identical because of the difference in the Memento-Datetime values.



Since PRONI used the same replay engine as NLI (when both were hosted by the European Archive/IMF), it returned the status code \texttt{200 OK} for archived \texttt{4xx/5xx} responses. For example, when requesting the memento \url{http://webarchive.proni.gov.uk/20160216154000/http://www.megalithic.co.uk/} on 2017-11-18, the original archive returned \texttt{200 OK} for an archived \texttt{403 Forbidden}.  Even the HTTP status code of the inner iframe in which the archived content is loaded returned \texttt{200 OK}. When requesting the corresponding memento \url{http://wayback.archive-it.org/11112/20160216154000/http://www.megalithic.co.uk/},  Archive-It properly returns the status code \texttt{403} for an archived \texttt{403} response.


Mementos may disappear when moving from the original archive to the new archive. For example, the request to the URI-M
\url{http://webarchive.proni.gov.uk/20140408185512/http://www.www126.com/}
from the original archive resulted in \texttt{200 OK}.
The request to the corresponding URI-M
\url{http://wayback.archive-it.org/11112/20140408185512/http://www.www126.com/}
from Archive-It results in \texttt{404 Not Found}.
Before transferring collections to the new archive, it is possible that the original archive reviews collections and removes URI-Rs/URI-Ms that are considered off-topic \cite{tpdl15:off-topic,ipres2018:off-topic,alnoamany-offtopic-ijdl} or spam (e.g., the URI-R \url{www.www126.com} is about auto insurance).  

\subsection{Perma.cc}

The Perma.cc archive \cite{zittrain2014perma} is maintained by the Harvard Law School Library and has the goal of providing permanent URIs for archived webpages for use in academic publications. When a user archives a webpage in Perma.cc, the user is provided a unique ``Perma Link" as the URI-M (e.g., \url{https://perma.cc/T8U2-994F}). This is different than the URI-Ms from many other archives that include the 14-digit Memento-Datetime and URI-R in the URI-M.  

Prior to 2020, mementos were accessible via long-form URI-Ms, for instance \url{http://perma-archives.org/warc/20170731024959/https://www.tmall.com/}. These long-form URI-Ms were what had been returned to the LANL Memento Aggregator and are the form we used in our longitudinal study.


On February 5, 2020, Perma.cc deployed new support for Memento, which involved changing the endpoints for Memento services and the URI-Ms provided  \cite{permacc-memento-update}.  Some of the changes included removing access to the URI-Ms of the form \url{http://perma-archives.org/warc/...} that we had been using and only returning mementos with the 
Perma Link URI-Ms, such as \url{https://perma.cc/T8U2-994F}.  Another change removed embedded resources from Memento access and began only providing access to top-level pages that were public, user-initiated captures.  Requests for mementos of non-top level pages would return \texttt{404 Not Found}.

\begin{figure}[ht!]
\centering
 \fbox{
\includegraphics[width=0.85\textwidth, trim =0 40 40 50, clip]{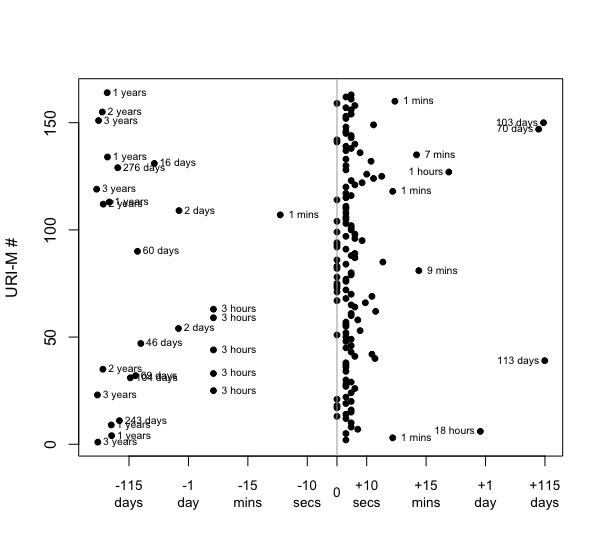}
}
\caption{Difference between the Memento-Datetimes for the long-form Perma.cc URI-Ms and the corresponding short-form URI-Ms.} 

\label{fig:perma}
\end{figure}

Our original study included 182 long-form URI-Ms from Perma.cc. After the change, we were able to find only 164 corresponding short-form URI-Ms, resulting in 18 mementos that could not be found at all. It is possible that these missing mementos were not top-level URI-Ms or that they were private Perma Links, both of which are no longer replayable.
However, in all 164 cases, the corresponding mementos had different Memento-Datetime values, with delta ranging from one second to three years. Figure \ref{fig:perma} shows each URI-M and the difference between the original Memento-Datetime and the new Memento-Datetime. 

\section{Conclusion}

The main goal of web archives is to provide permanent access to web resources using consistent URIs. Links to such archived resources are used in academic publications so that the information cited remains available even if the resource on the live Web changes or disappears. Our study provides a cautionary tale for archives that have to change domains
or web archiving platforms. In a study of 16,627 mementos over 17 public web archives, we found that four archives changed their domains without providing a machine-readable notification, affecting 1,981 mementos from our study.  Of these, we were not able to find 537 identical mementos in the new archives, 20 of which had disappeared completely. The data set is available in GitHub \cite{maturban-dataset}.

%
%
\bibliographystyle{splncs04}
\bibliography{refs}

\end{document}